\documentclass{epl}

\title{Unbinding Transition Induced by Osmotic Pressure in Relation to
Unilamellar Vesicle Formation}

\author{N. L. Yamada\inst{1}\thanks{E-mail: \email{yamadan@post.kek.jp}}
\and M. Hishida\inst{2} \and H. Seto\inst{2} \and K. Tsumoto\inst{3}
\and T. Yoshimura\inst{3}}
\institute{
  \inst{1} Institute of Materials Structure Science, High Energy
  Accelerator Research Organization -
  Oho, Tsukuba, Ibaraki, 305-0801, Japan\\
  \inst{2} Department of Physics, Kyoto University -
  Kitashirakawa-oiwake-cho, Sakyo, Kyoto, 606-8502, Japan\\
  \inst{3} Department of Chemistry for Materials, Mie University -
  Kurimamachiya-cho, Tsu, Mie, 514-8507, Japan
}
\shorttitle{Unbinding Transition Induced by Osmotic Pressure}
\shortauthor{N. L. Yamada et al.}


\begin{document}

\maketitle

\begin{abstract}
 Small-angle X-ray scattering and phase-contrast microscopy experiments
 were performed to investigate the effect of the osmotic pressure on
 vesicle formation in a dioleoylphosphatidylcholine (DOPC)/water/NaI
 system.
 Multi-lamellar vesicles were formed when a pure lipid film was hydrated
 with an aqueous solution of NaI.
 On the other hand, uni-lamellar vesicles (ULVs) were formed when a
 lipid film mixed with an enough amount of NaI was hydrated.
 To confirm the effect of the osmotic pressure due to NaI, a free-energy
 calculation was performed.
 This result showed that the osmotic pressure induced an unbinding
 transition on the hydration process, which resulted in ULV formation.
\end{abstract}

\section{Introduction}
All biomembranes mainly consist of lipid bilayers, and functional
proteins float on them.
Such bilayers also appear in aqueous solutions of synthesized
phospholipids.
Therefore, bilayers in such solutions are intensively studied to
understand the physical properties of biomembranes.
Moreover, these bilayers form vesicles that attract attention for
studying model cells.
Such vesicles can be effectively obtained by hydrating dry lipid films
on substrates or test tubes. (natural swelling method \cite{Reeves})
For example, dioleoylphosphatidylcholine (DOPC) films, one of the
typical phospholipids in biomembranes, provide micrometer-size vesicles.
These DOPC vesicles normally grow with a multi-bilayer shell
(multi-lamellar vesicle; MLV), although uni-lamellar vesicles (ULVs) are
prefered for using as model cells.
Therefore, effective methods to create ULVs have been proposed
\cite{Darszon,Kim,Lasic,Angelova,Magome,Akashi,Angelova2,Yamashita}.

One of the authors has recently designed a novel method for preparing
ULVs by hydrating dry lipid films mixed with sugar and salt
\cite{Tsumoto}.
As some previous studies have pointed out, an osmotic pressure would
promote ULV formation
\cite{Magome,Angelova2}.
The mechanism of ULV formation induced by osmotic pressure, however, has
not been yet clarified.

A key phenomenon to understand the mechanism of ULV formation is an
``unbinding transition'', in which the inter-bilayer distance diverges
infinitely and the bilayer stacking is unstabilized.
First, this transition was predicted theoretically by Lipowsky
\cite{Lipowsky}.
Recently, some experiments on this subject have been performed using
small-angle X-ray scattering, small angle neutron scattering, and X-ray
reflectivity
\cite{Vogel,Deme,PozoNavas,Yamada}.
The authors recently clarified the origin of the unbinding transition in
terms of the interaction between bilayers;
a collaboration of the long- and short-range repulsive force, for
example a steric repulsion due to membrane undulation \cite{Helfrich}
and electrostatic repulsion, would be the origin of this transition
\cite{Yamada}.
Therefore, the osmotic pressure would play an important role for the
unbinding transition as these repulsions.

In this study, we prepared two kinds of samples to investigate the
effect of the osmotic pressure when dry lipid films include an additive
before hydration.
One is an ``NaI in film'' sample in which a dry DOPC film mixed
with NaI was hydrated with pure water, that is, an ``NaI in film''
sample is made by using the novel method to create ULVs.
The other is  an ``NaI in solution'' sample in which a dry DOPC film was
hydrated with an aqueous solution of NaI for a control experiment.
In this condition, DOPC bilayers were in the liquid-crystalline phase at
room temperature, and vesicles were effectively formed by hydration
\cite{Hishida}.
NaI is suitable to prepare a homogeneous lipid film for ``NaI in film'',
since it is a good solute for both water and methanol.
(The details are described in the Experimental section.)
To investigate the dependence of the vesicle formation on the sample
preparation, the structure of the multi-bilayer shell were observed
using small-angle X-ray scattering (SAXS); also, the shape of vesicles
were observed by phase-contrast microscopy (PCM).

\section{Experiments}
DOPC was purchased from Sigma Chemical Inc. and NaI from Wako Pure
Chemical Industries Ltd.
To prepare dry lipid films of ``NaI in solution'' samples, DOPC
solutions of organic solvents, 1:2(v/v) methanol/chloroform, were
evaporated under N$_2$ gas flow in a glass test tube, and kept in a
vacuum at room temperature overnight.
The obtained lipid films were hydrated with an NaI aqueous solution with
various concentrations of NaI.
For ``NaI in film'' samples, methanol solutions of various NaI
concentrations were prepared before mixing with chloroform.
Then, DOPC was dissolved in these organic solvents, the organic solvents
were evaporated in a vacuum as the ``NaI in solution'' samples, and the
obtained DOPC/NaI films were hydrated in pure water.

The SAXS experiments were performed at the BL40B2 beam port of SPring8
at Japan Synchrotron Radiation Research Institute (JASRI).
The incident X-rays were monochromatized by a double-crystal
monochromator, and the wavelength was 1~\AA\ ($\Delta E/E \simeq
10^{-4}$).
The detector was an imaging-plate area detector placed at 1~m from the
sample position.
The samples were prepared to be 1wt.\% of DOPC concentration, and to
have the molar fractions of DOPC:NaI=10000:1, 1000:1, 100:1, 10:1, and
1:1.
Since all of the obtained two-dimensional data had no preferred
orientation, they were azimuthally averaged to provide one-dimensional
data.
All of the experiments were performed at room temperature.

The PCM experiments were performed by using Nikon TE-300 and recorded on
S-VHS videotape at 30~frames/sec.
The DOPC concentration was 1~mM (about 0.08wt.\%) so as to avoid vesicle
aggregation, and the molar ratio of DOPC:NaI was 1:1.
The PCM experiments were also performed at room temperature as SAXS
experiments.

\section{Result}
\begin{figure}[tbp]
 \begin{center}
  \includegraphics[width=\textwidth,keepaspectratio]{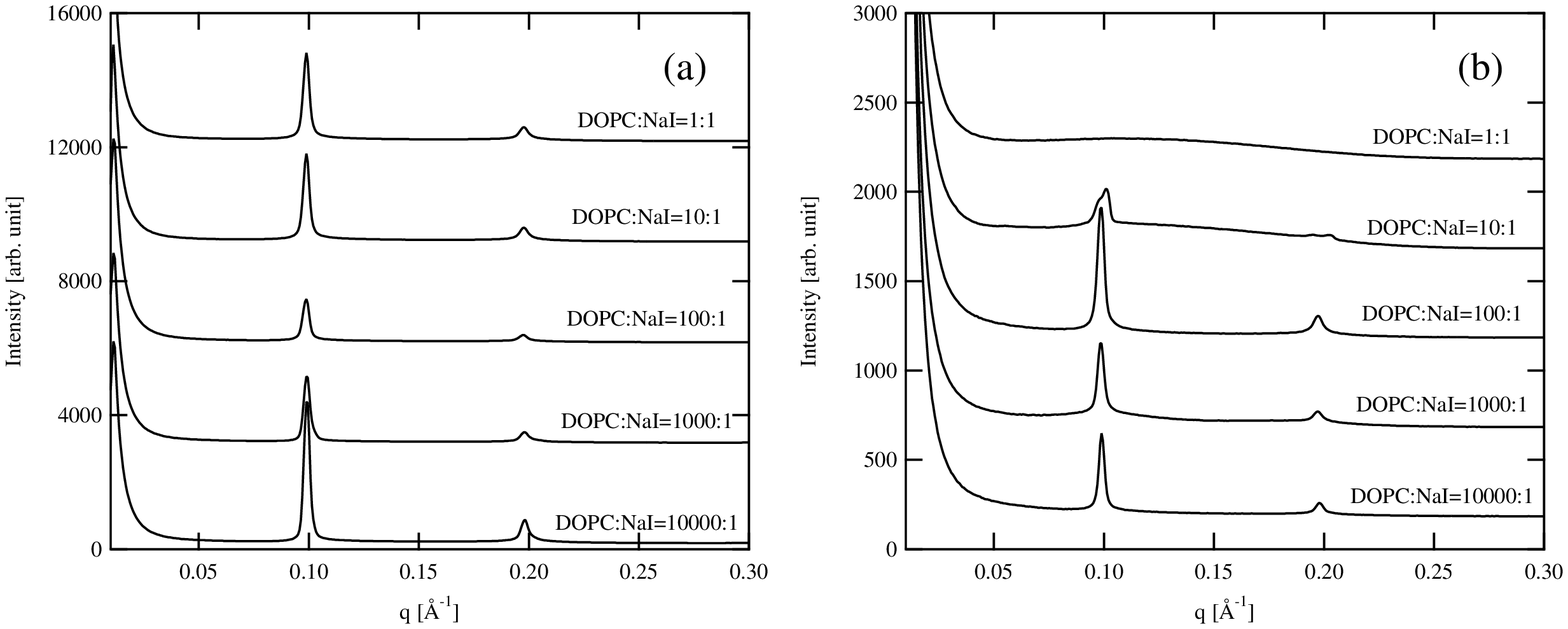}
  \caption{Dependence of SAXS profiles on the molar ratio of NaI to
  DOPC.
  (a) SAXS profiles of ``NaI in solution'' samples.
  The obtained profiles are independent of the molar ratio.
  (b) SAXS profiles of ``NaI in film'' samples.
  The obtained profiles drastically change above DOPC:NaI=10:1.
  The profiles of higher NaI ratio were shifted for better
  visualization.}
  \label{fig:SAXS_all}
 \end{center}
\end{figure}
Figure~\ref{fig:SAXS_all}(a) shows the SAXS profiles obtained from the
``NaI in solution'' samples.
All profiles have sharp Bragg peaks due to the regular stacking of lipid
bilayers, whose repeat distance, $d$, is 63.5~{\AA}.
This value is almost the same as that of an aqueous solution of DOPC,
$d=63.1$~\AA
\cite{STNagle}.
This means that NaI molecules in water had no effect on the lamellar
structure.
Figure~\ref{fig:SAXS_all}(b) shows the SAXS profiles obtained from the
``NaI in film'' samples.
Although the SAXS profiles were essentially the same as the ``NaI in
solution'' samples at lower NaI concentration, the Bragg peaks split
into two peaks at DOPC:NaI=10:1, and completely disappeared at
DOPC:NaI=1:1.
This evidence imply that the osmotic pressure due to NaI between
bilayers destroys the multi-lamellar structure.

\begin{figure}[tbp]
 \begin{center}
  \includegraphics[width=\textwidth,keepaspectratio]{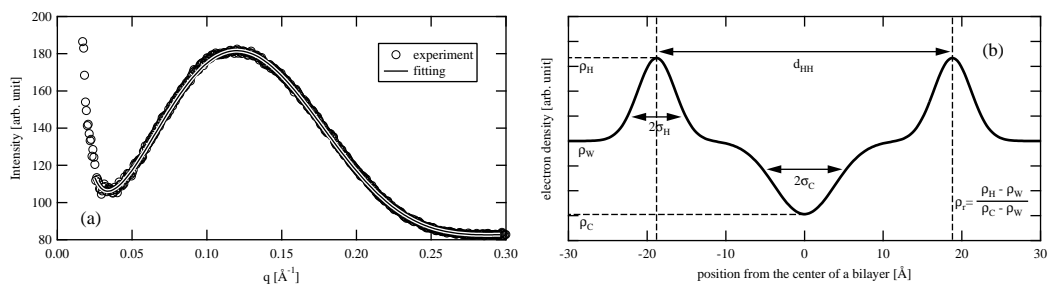}
  \caption{(a) Magnified view of the SAXS profile from the ``NaI in
  film'' sample of DOPC:NaI=1:1 with the result of fitting.
  (The same as the uppermost profile of Fig.~\ref{fig:SAXS_all}(b).)
  The fitting model reproduced the experimental result well.
  (b) Electron density profile obtained from the fitting.
  $\sigma_H$ is the standard deviation of the electron density of head
  groups, $\sigma_C$ the standard deviation of the electron density of
  an end of hydrocarbon tails, $\rho_r$ the ratio of the peaks at the
  head groups to that at the end of hydrocarbon tails, and $d_{HH}$ the
  head-to-head distance of a bilayer.}
  \label{fig:SAXS_Fq}
 \end{center}
\end{figure}

To confirm that lipid bilayers still exist in the sample of
DOPC:NaI=1:1, we carried out a model fitting for the obtained SAXS
profile.
Since the structure factor should be unity because of no correlation
between the bilayers, the fitting was performed using
\begin{equation}
 I(q)=I_\rho(q)+\frac{|F(q)|^2}{q^2}+B,
\end{equation}
where $F(q)$ is the form factor of the bilayers, $I_\rho(q)$ the
scattering due to the concentration fluctuation of the lipids, and $B$
the constant background.
In this model, $I_\rho(q)$ was assumed to be Lorentzian \cite{Nallet},
\begin{equation}
 I_\rho(q)\propto\frac{1}{q^2\xi^2+1},
\end{equation}
where $\xi$ is the correlation length for the concentration fluctuation.
$F(q)$ was assumed to be the Fourier transform of three Gaussians, which
represents the bilayer electron density, as the following equation
\cite{Pabst}:
\begin{equation}
 F(q)\propto2\sigma_H\exp\left[-\frac{(\sigma_Hq)^2}{2}\right]
  \cos\left[\frac{d_{HH}q}{2}\right]
  +\rho_r\sigma_C\exp\left[-\frac{(\sigma_Cq)^2}{2}\right].
\end{equation}
Here, $\sigma_H$, $\sigma_C$, $\rho_r$, and $d_{HH}$ are the parameters
used to describe the electron density profiles of the lipid bilayers, as
shown in fig.~\ref{fig:SAXS_Fq}(b).
The fitting was performed with the least-square method for a $q$ range
of $0.025<q<0.3$~[\AA$^{-1}$].

As shown in fig.~\ref{fig:SAXS_Fq}(a), a broad peak corresponding to
the correlation between the head groups appeared around
$q=0.12$~\AA$^{-1}$.
The fitting function reproduced the experimental result well, and the
fitting parameters were evaluated to be $\sigma_H$=2.38~{\AA},
$\sigma_C$=3.85~{\AA}, $\rho_r$=-1.13, and $d_{HH}$=37.6~{\AA}, which
agrees with $d_{HH}=35.3$~{\AA} from the literature
\cite{STNagle}.
(The electron density profile calculated from these parameters is shown
in fig.~\ref{fig:SAXS_Fq}(b).)
Since the electron density of the phosphate group is much greater than
that of the other groups in the DOPC bilayers, $d_{HH}$ is the most
characteristic parameter in the electron density profile.
Therefore, the agreement of its value confirms that only the scattering
due to the form factor was seen in the profile from the ``NaI in film''
sample at DOPC:NaI=1:1.
This suggests that the unbinding transition would occur in the ``NaI in
film'' sample, since the Bragg peak due to the correlation between
bilayers disappeared.

\begin{figure}[tbp]
 \begin{center}
  \includegraphics[width=0.7\textwidth,keepaspectratio]{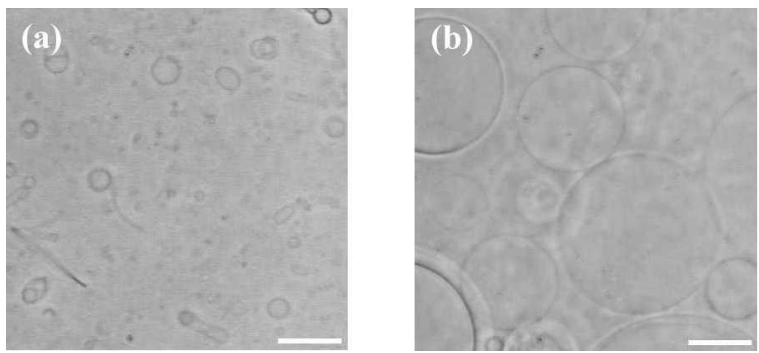}
  \caption{Difference of the phase contrast images of the vesicles
  obtained by the two preparation methods at molar ratio of
  DOPC:NaI=1:1.
  (a) Vesicles in the ``NaI in solution'' sample.
  The size of the vesicles was below a few $\mu$m and their shells were
  thick.
  (b) Vesicles in the ``NaI in film'' sample.
  The size of the vesicles was over 10~$\mu$m and their shell
  thicknesses were thin.
  The scale bar is 10~$\mu$m.}
  \label{fig:PCM}
 \end{center}
\end{figure}
Figure~\ref{fig:PCM} shows the vesicles observed by PCM.
The difference between the preparation procedures was clear: smaller
vesicles with a thick shell formed in the ``NaI in solution'' sample,
whereas larger vesicles with a thin shell formed in the ``NaI in film''
sample.
This is consistent with the disappearance of the Bragg peak in the ``NaI
in film'' sample.
Therefore, we concluded that an addition of NaI to DOPC films before
hydration promotes ULV formation
\cite{Tsumoto}.

\section{Discussion}
\begin{figure}[tbp]
 \begin{center}
  \includegraphics[height=5cm,keepaspectratio]{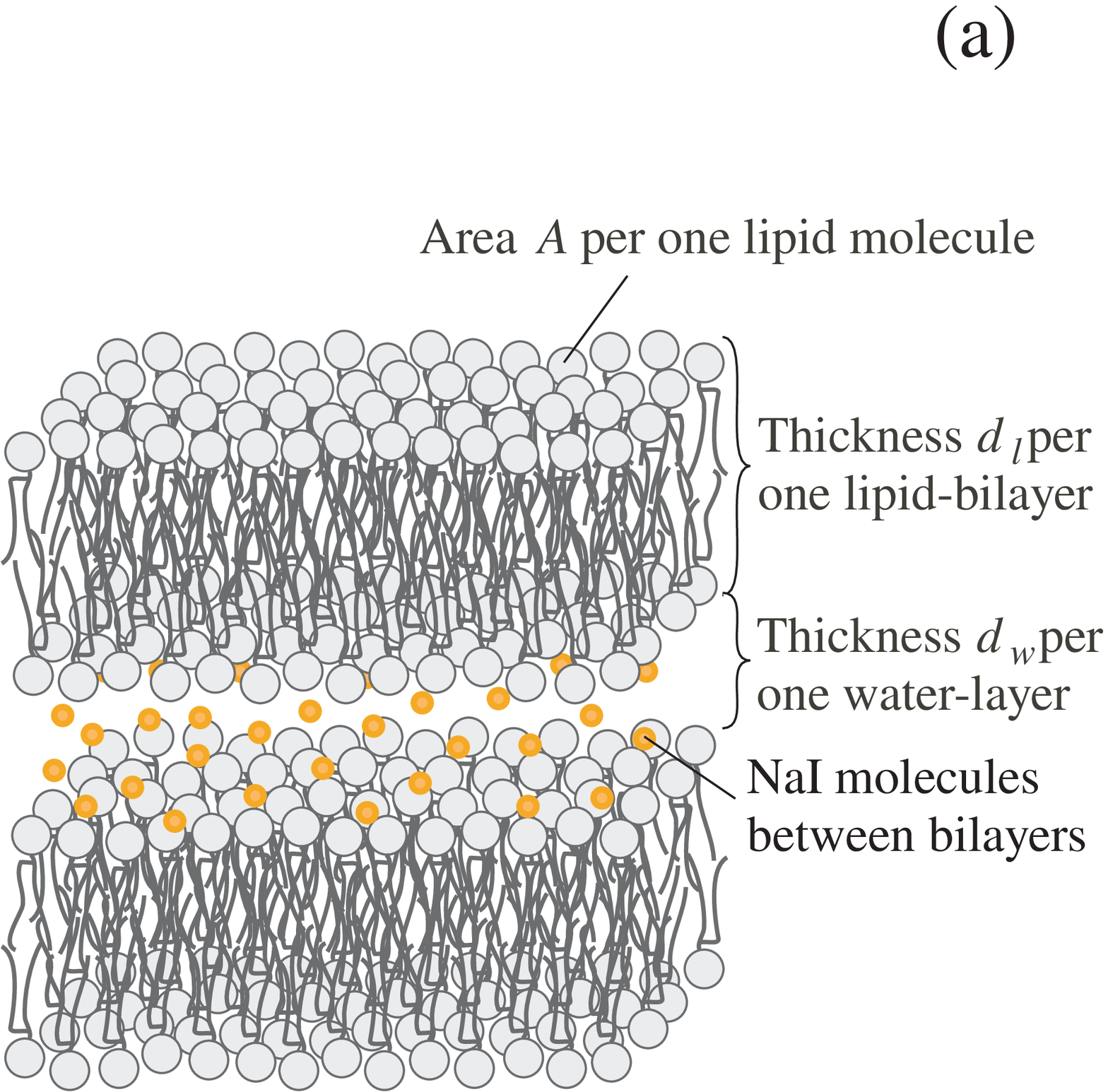}
  \hspace*{1cm}
  \includegraphics[height=5cm,keepaspectratio]{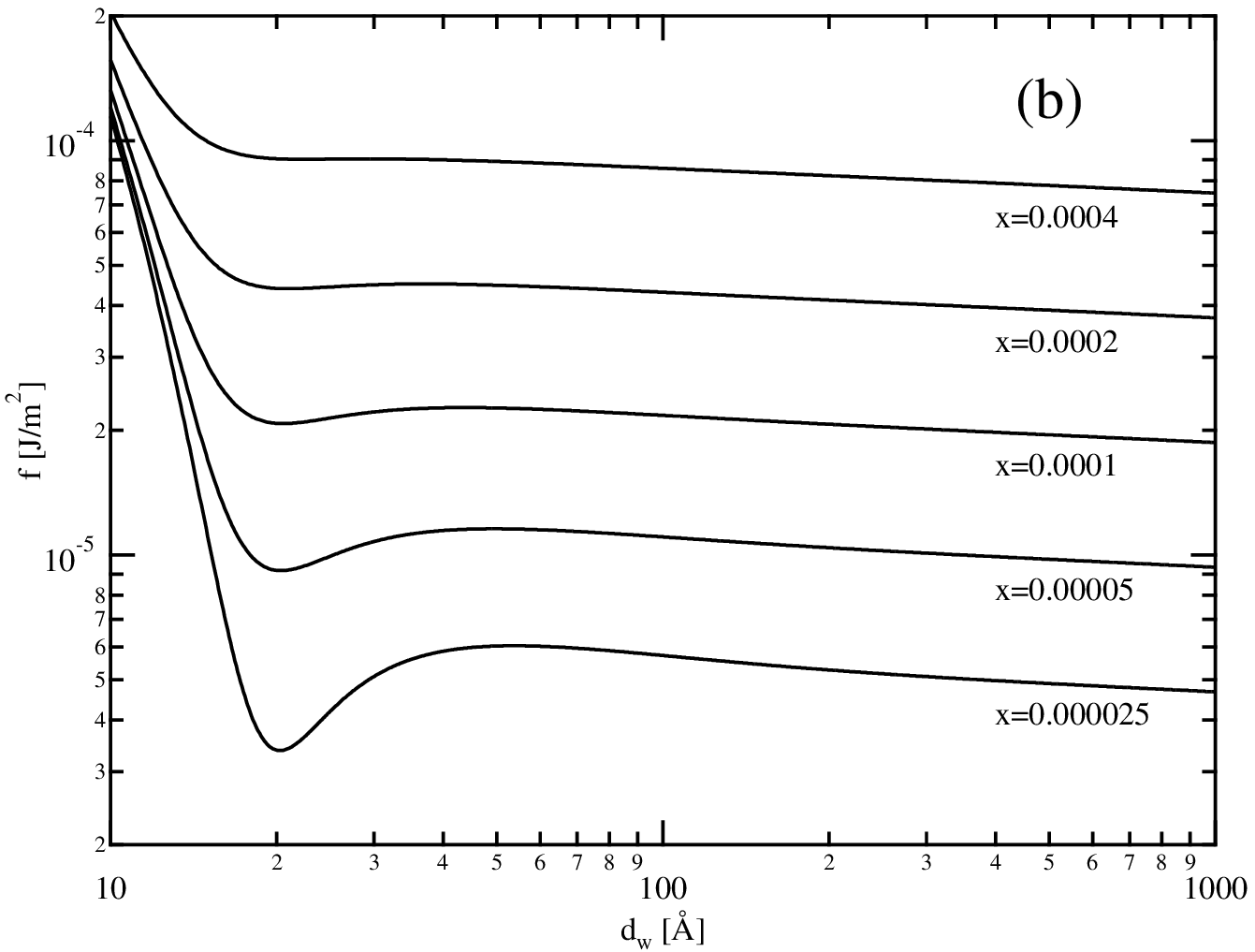}
  \caption{(a) Schematic illustration of the assumption in the
  free-energy calculation
  \cite{Yamada2}.
  There are two parallel and flat bilayers and NaI molecules located
  between them.
  (b) Calculated free energy densities with changing the molar
  ratio of NaI, $x$.
  The shift to upside with increasing $x$ is not due to an offset, but
  to the change of $f_{osm}$.}
  \label{fig:calc_f}
 \end{center}
\end{figure}

To discuss the relation between the unbinding transition and the ULV
formation in the ``NaI in film'' sample, we calculated the free-energy
density profile of the outermost bilayer depending on the inter-bilayer
distance.
In this calculation, we assumed the following situation.
First, a lipid bilayer is known to stack on the surface of a substrate
before hydration.
It is reasonable to assume that NaI molecules in a dry lipid film locate
between the bilayers, as shown in fig.~\ref{fig:calc_f}(a)
\cite{Yamada2}.
Next, only water can penetrate through the lipid bilayers on the
hydration process, since the bilayer is known to be semipermeable.
Furthermore, the free-energy density caused by osmotic pressure,
$f_{osm}$, arises from the concentration difference of NaI between
inside and outside of the bilayers
\cite{f_osm}.
Finally, the inter-bilayer distance increases up to a stable distance,
where the free-energy profile has a local minimum.
According to previous studies, three other considerable free-energy
densities should be considered: the first one is due to the van der
Waals attractive force, $f_{vdW}$, the second one is due to the
hydration layers, $f_{hyd}$, and the third one is from the steric
repulsion due to the membrane undulation, $f_{st}$
\cite{STNagle,Yamada}.
Therefore, the total free energy density $f$ can be described as
follows:
\begin{eqnarray}
 f&=&f_{vdW}+f_{hyd}+f_{st}+f_{osm},\\
 f_{vdW}&=&-\frac{H}{12\pi}\left\{
  \frac{1}{d_w^2}-\frac{2}{(d_w+d_l)^2}+\frac{1}{(d_w+2d_l)^2}\right\},\\
 f_{hyd}&=&P_h\lambda\exp\left[-\frac{d_w}{\lambda}\right],\\
 f_{st}&=&0.42\frac{(k_BT)^2}{K_cd_w^2},\\
 f_{osm}&=&-\frac{2xk_BT}{A}\ln[d_w],\label{eq:f_osm}
\end{eqnarray}
where, $H$ is the Hamaker constant, $d_w$ the inter-bilayer distance,
$d_l$ the bilayer thickness, $P_h$ the prefactor of $f_{hyd}$, $\lambda$
the decay length of hydration layers, $K_c$ the bending rigidity of
bilayers, $x$ the molar ratio of NaI to DOPC, and $A$ the area per DOPC
molecule.

\begin{table}[tbp]
 \begin{center}
 \caption{Parameters determined by Tristram-Nagle
  \cite{STNagle}}.
  \begin{tabular}{ccccccc}
   \hline
   $T$ & $d_l$ & $A$ & $K_c$ & $H$ & $P_h$ & $\lambda$ \\ \hline
   30$^\circ$C & 45.3 \AA & 72.2 \AA$^2$ & $1.0\times10^{-19}$ J &
   $4.0\times10^{-21}$ J & $5.0\times10^{7}$ J/m$^3$ & 2.26 \AA
   \\ \hline
  \end{tabular}\label{tab:constant}
 \end{center}
\end{table}
Figure~\ref{fig:calc_f}(b) shows the calculated free-energy densities as
a function of $d_w$ by using these expressions with the parameters
determined by Tristram-Nagle, as shown in table~\ref{tab:constant}.
A local minimum is seen at $d_w=20.3$~{\AA} in the free-energy profile
of the lowest NaI content ($x=0.000025$).
The obtained $d$ value ($d=d_l+d_w=65.6$~\AA) agrees with that
obtained from the SAXS profiles ($d=$63.5~\AA).
Therefore, this local minimum corresponds to the inter-bilayer distance
of the bilayer stacking.
With increasing the molar ratio of NaI, this local minimum becomes
shallow and disappears above $x=0.0004$ due to an increase of the
osmotic pressure.
This results in an increase of the inter-bilayer distance up to
infinitely, {\it i.e.}, the unbinding transition takes place.
Therefore, we concluded that the osmotic pressure due to the additive
induced the unbinding transition on the hydration, which resulted in the
ULV formation in the ``NaI in film'' samples.

It should be noted that the calculated critical concentrations to induce
the unbinding transition were different from the SAXS result; more than
10\% of NaI against DOPC was required to induce the unbinding
transition, while only about 0.04\% of NaI was required in the model
calculation.
This would have originated from an over-estimation of the osmotic
pressure, since the NaI molecules should locate not only between
bilayers, but also at the outside of bilayers in the multi-layer lipid
film.
This problem is, however, not essential with respect to the mechanism of
vesicle formation.

\section{Conclusion}
A difference of structure, depending on the process of mixing lipids
with salt and hydration (``NaI in solution'' and ``NaI in film''), was
investigated with changing the molar ratio of NaI to DOPC using SAXS and
PCM.
The SAXS experiment suggested that a large amount of NaI in the ``NaI in
film'' sample induced the unbinding transition, and the PCM experiment
showed that NaI in the ``NaI in film'' promoted the formation of ULVs.
The calculation of the free energy for the ``NaI in film'' sample was
also performed, and showed that the osmotic pressure was the origin of
the unbinding transition.
From these results, we concluded that the osmotic pressure due to much
amount of NaI in the lipid film induces an unbinding transition which
accelerates the formation of ULVs.

\acknowledgements
 SAXS experiments were carried at BL40B2 of SPring-8 with the
 approval of the Japan Synchrotron Radiation Research Institute (JASRI).
 (Proposal No. 2004B0520-NL2b-np)
 We are grateful to Dr. K. Inoue and Dr. S. Sasaki (JASRI) for
 experimental supports.
 One of the authors (H.~S) was supported by a Grant-in-Aid for
 Scientific Research (No. 17540382), by a Grant-in-Aid for the 21st
 Century COE ``Center for Diversity and Universality in Physics'' from
 the Ministry of Education, Culture, Sports, Science, and Technology
 (MEXT) of Japan, and by the Yamada Science Foundation.


\begin{thebibliography}{}
\bibitem{f_osm} The osmotic pressure $p_{osm}$ due to the number of the
	molecule $N$ per volume $V$ is $Nk_BT/V$.
	In this case, $N$ and $V$ in a unit cell are $2x$ and $Ad_w$,
	respectively.
	On the other hand, the relation between $f_{osm}$ and $p_{osm}$
	is described as $p_{osm}=-\partial f_{osm}/\partial d_w$
	from their definition.
	Therefore, we obtain eq.~(\ref{eq:f_osm}) under the assumption
	that any NaI molecules are not ionized.
\end{thebibliography}
\end{document}